# Hybrid Forecasting of Exchange Rate by Using Chaos Wavelet SVM-Markov Model and Grey Relation Degree

Kim Gol [a]   Ri Suk Yun [b]

[a] Center of Natural Sciences, University of Sciences, DPR Korea
[b] Foreign Economic General Bureau, Pyongyang, DPR Korea

5 June, 2012

**Abstract**

This paper proposes an exchange rate forecasting method by using the grey relative combination approach of chaos wavelet SVM-Markov model. The problem of short-term forecast of exchange rate by using the comprehensive method of the phase space reconstitution and SVM method has been researched

We have suggested a wavelet-SVR-Markov forecasting model to predict the finance time series and demonstrated that can more improve the forecasting performance by the rational combination of the forecast results through various combinational tests.

Our test result has been showed that the two-stage combination model is more excellent than the normal combination model. Also we have comprehensively estimated the combination forecast methods according to the forecasting performance indicators.

The estimated result have been shown that the combination forecast methods on the basic of the degree of grey relation and the optimal grey relation combination have fine forecast performance.

## Introduction

The time series analysis is importantly suggested on the many application including the control of physical systems, process of engineering, biochemistry, environmental economic system, company management and economy. Especially, the forecast of exchange rate acts on the macroeconomic policy, business management and individual decision-making establishment.

Owing to these importances, the forecast of exchange rate became a hot spot of research for many scholars in the world. There are many methods in forecast technique of time series such as linear regression, Kalman filtering, neural network modeling, and fuzzy system method.

The linear regression method is simple but its adaptation is poor. The filtering method has the adaptation but it is essentially linear technique.

The neural network modeling method approximates arbitrary nonlinear function but it demands many learning data and difficult to analysis.

With contrast for it, fuzzy system method has the good explanatory ability but its adaptation is poor.

The support vector machine(SVM) has effectively been used for machine learning.

Its reason is owing to that the generalized properties of SVM are not dependent on the completely many training data but it is only dependent on the subset of training data namely the support vectors.

At present time, SVM are widely used in multi-field such as handwriting recognition, three dimensions object recognition, regression analysis and so on.

Recently, SVM is used as the universal tool in prediction of time series. ([1-13])

Its reason is owing to that the SVM has the considerable feature which it has the generalized performance and it don't fall into the local minimization and it is represented with few information data.

The training error has been minimized in the mostly traditional forecast technique which implemented the principle of empirical danger minimization. With contrast for it, SVM is implement ting the principle of structural danger minimization to find the upper bound and lower bound of generalized error.

This is immediately the reason which SVM become to be better for the forecast performance in the generalized performance than other traditional methods.

SVM have always the global optimum solution and don't fall into the local minimization because of its learning is equivalent to solve the convex quadratic programming problem with linear constrain.

Actually, its solution is only to be determined by the support vectors which are subsets of training data set. Therefore, it is easily capable of notating.

The first importance step to enhance the forecast performance of SVM is the feature selection or feature extraction of time series.

We have made use of the embedding dimension([40]) of chaos time series in order to characterize the input sample number needed for the machine learning of SVM.

Consequently, we have researched the problem of short-term forecast of exchange rate by using the comprehensive method of the phase space reconstitution([41]) and SVM method.

We have suggested a Wavelet-SVR-Markov forecasting model to predict the finance time series and demonstrated that can more improve the forecasting performance by the rational combination of the forecast result through various combinational tests.

Our test result has been showed that the two-stage combination model is more excellent than the normal combination model.

Also we have comprehensively estimated the combination forecast methods according to the forecasting performance indicators.

The estimated result have been shown that the combination forecast methods on the basic of the degree of grey relation and the optimal grey relation combination have fine forecast performance.

# 1. The phase space reconstitution
## 1.1. The theory of phase space reconstitution of the chaos time series

Recently, the analysis of nonlinear system especially time series generated under the chaos background becomes to regard as important more and more for person.

The chaos phenomenon which there exists widely in the natural world is a kind of irregular motion and it is one of the complicated behaviors generated by the deterministic nonlinear dynamic system. The analysis of chaos time series is an important application field. It not only can be progressed the detection and diagnosis of chaos through the determination of dynamic model for the system which are going to research but also it can be able to apply widely in the various field of the natural science and social science.

It has the very important signification in theory and practice, for example, the forecasting of stock market price and price fluctuation, the forecasting of glacial epoch.

Recently, according to the intensification of the research for chaos theory and applied technology constantly, the model-building and the forecasting of chaos time series becomes to a hot spot research field in the forecasting of stock and exchange rate.

Until now, the persons have already suggested various forecast method of chaos series on the basic of Takens embedding theorem and the idea of phase space reconstruction.

The foundation of the forecast of chaos time series is the theory of phase space reconstruction. Through the phase space reconstruction, one can find the law of evolution which is behinded in chaotic attractor. And, by entering the present data to range which can be described, one can obtain a novel method and idea to research the time series.

[Definition 1] Let $(N_1, \rho_1)$ and $(N_2, \rho_2)$ are tow metric space.

We assume that there exists mapping $\varphi: N_1 \to N_2$ and it holds true the following conditions:

（ⅰ） $\varphi$ is surjective mapping

（ⅱ） $\rho_1(x, y) = \rho_2(\varphi(x), \varphi(y))$, $\forall x, y \in N$

Then $(N_1, \rho_1)$ and $(N_2, \rho_2)$ is called an isometric isomorphism.

[Definition 2] Let $(N_1, \rho_1)$ and $(N_2, \rho_2)$ are tow metric space.

Let $(N_0, \rho_0)$ is a subspace of metric space $(N_2, \rho_2)$.

If $(N_1, \rho_1)$ and $(N_0, \rho_0)$ is called an isometric isomorphism, then it is called that metric space $(N_1, \rho_1)$ can be embedded to metric space $(N_2, \rho_2)$.

*[Takens theorem]*

Let $M$ be a $d$ dimensional manifold and $\varphi: M \to M$ be a smooth diffeomorphism mapping

And let $\gamma: M \to R$ be a second order continuously differentiable mapping and $\phi(\varphi, \gamma)$ be a mapping $\phi(\varphi, \gamma)$ $M \to M^{2d+1}$. Here $\phi(\varphi, \gamma) = (\gamma(x), \gamma(\varphi(x)), \gamma(\varphi^2(x)), \cdots, \gamma(\varphi^{2d}(x)))$.

Then, $\phi(\varphi, \gamma)$ is a embedding from $M$ to $R^{2d+1}$

For a time series $\{x_1, x_2, \cdots, x_{n-1}, x_n, \cdots\}$, if we can suitably choice the embedding dimension $m$ and delay time $\tau$, then can reconstruct phase space:

$$Y(t_i) = [x(t_i), x(t_i + \tau), x(t_i + 2\tau), \cdots, x(t_i + (m-1)\tau)] \quad (i = 1, 2, \cdots) \tag{1}$$

By Takens theorem, we can immediately return to the dynamic character of the attractor under the meaning of phase equivalence.

The theoretical dependence of chaos forecast for finance time series is the delay embedding theorem subjected by Takens and Mance.

We assume that an actual data $\{y(k)\}$ $(k=1,2,\cdots,N)$ is output of certain state for one dimensional system.

By using the delay embedding theorem, the time series $\{y(k)\}$ of one dimensional dynamic system has been embedded to $D_E$-dimensional space( the embedding delay time is $\tau$ ) that is

$$y(k) \Rightarrow \bar{y}_R(k) = [y(k), y(k-\tau), y(k-2\tau), \cdots, y(k-(D_E-1)\tau)]$$

Let "$\{D_E, \tau\}$" be a window with which impute dimension is $D_E$ and delay time is $\tau$.

This window has made to appear all elements of $D_E$-dimension among time series $\{y(k)\}$ at the same time. When the time series have been moved sequentially and passed this window, we can obtain $D_E$-dimension state vector such as:

$$Y_k = [y(k), y(k-\tau), y(k-2\tau), \cdots, y(k-(D_E-1)\tau)]^T, k=1,2,\cdots,N_m \quad (2)$$

Here $N_m = N - (D_E-1)\tau$ is number of reconstructed vectors.

The method which the state vectors have been obtained by such method from the time series $y(k)$ is called an embedding methods.

The space composted from the vector $Y_k$ is called a quasi-phase space.

Let $N$ be a dimension of original phase space. If condition $m \geq 2N+1$ holds true, then Takens theorem [2] shows that the quasi-phase space and the phase space of systemic is diffeomorphism.

That is, above two spaces is the equivalent space topologically.

Therefore, those have perfectly same dynamic character each other.

**1.2. The determination of embedding dimension and delay time**

One of the crucial problems of reconstruction for phase space is to determine the delay time $\tau$.

There are many methods to choose of $\tau$ for actual problem already.

The method often used is the self-correlation function method and the method of mutual amount of information.

By using method of average mutual amount of information, we have determined the embedding dimension.

The delay is choused as $\tau$ which average mutual information $I(\tau)$ minimize and the embedding dimension is determined by the procedure of false nearest neighbors[41].

The procedure of average mutual information method is given as follows.

When the time series $\{y(k)\}$ is given, we divide the change width of $\{y(k)\}$ to some subinterval (for example, subinterval of $B$ number)

We find out the probability which fall to the subinterval by number of average mutual information entering to each subinterval.

For the delay sequence $P(y(k-\tau))$, same procedure is too progressed.

Then, we compose the state set of $B^2$ number by the change width of sequence $\{y(k)\}$ between delay sequence $P(y(k-\tau))$ and find out joint probability $P(y(k), y(k-\tau))$ by number which $y(k)$ and $y(k-\tau)$ belong to each state set at the same time.

The average mutual information $I(\tau)$ is given by

$$I(\tau) = \sum_{y(k), y(k-\tau)} P(y(k), y(k-\tau)) \log_2 P(y(k), y(k-\tau))/(P(y(k))P(y(k-\tau)))$$

The procedure of false nearest neighbors is given as follows.

When the time series $\{y(k)\}$ is given, making increase the $D_E$ from 1 one by one, we find out the nearest neighbors of each point $Y_n = [y(n), y(n-\tau), \cdots y(n-(D_E-1)\tau)], n = 1, 2, \cdots, N-\tau$ in the $D_E$-dimensional space.

The Euclidean metric is used in the calculation of neighbors.

Then, adding component $y(n - D_E \tau)$ to $D_E$-dimension vectors, we judge whether the nearest neighbors of $D_E$-dimension stand also still to the false nearest neighbors in $D_E + 1$-dimension.

When incrementing the $D_E$, the number of the false nearest neighbors which is not done the neighbors in $D_E + 1$-dimension decreases as function of $D_E$.

When dropping the ratio of the false nearest neighbors to certain threshold value for instance below-1%, we choose its corresponding dimension as embedding dimension

In the case of our example, delay time and embedding dimension found out from above procedure are $\tau = 1$, $D_E = 4$ respectively

By this method, we have determined delay time and embedding dimension for SVR regression.

## 2. The smoothing of data and wavelet transformation

In the chaos forecasting for finance time series, by using delay and embedding theorem, one-dimension finance time series $\{y(k)\}$ is embed to $D_E$-dimension space (embedding delay time is $\tau$) and by using the state vector reconstructed, the forecasting for observable quantity is progressed

Actually, this problem is to approximate a mapping $f$ satisfying as following condition;

$$f : \mathrm{R}^{D_E} \to R$$
$$y(k+1) = f(\bar{y}_R(k)) \tag{3}$$

In the finance time series analyses, this is a non-well posed problem.

If the following three conditions that is, existence, uniqueness, continuity holds true at the same time, then given problem is called a well posed problem in meaning of Hadamard.

If it doesn't hold true, it is called a non-well posed problem,.

The cause occurred the non-well posed problem in dynamic system composed by finance time series is due to the following facts:

The first, the existence condition doesn't satisfied because of various unknown cause

The second, it is owing to the finance time series can't hold the sufficient information which is reflecting the dynamic situation of financial market.

Therefore, uniquely reconstructing the phase space for this system is impossible

The third, owing to the finance time series suffers the effect of various sudden accidents and hearsay, the fluctuation of system make strongly destroy the continuity condition.

Therefore, it is caused any uncertainty in the dynamic reconstruction

The basic method which non-well posed problem converts to well pose problem is to give the restriction for mapping $f : \mathrm{R}^{D_E} \to R$ by introducing various a priori knowledge.

The regularization theory of Tikhonov [8] is an effective method to solve this problem.

Now, the capacity of the financial market of some countries is comparatively small and law is incomplete, development doesn't maturate and fluctuation is excessive.

Therefore, necessarily one hand the approximation requirement for mapping $f : R^{D_E} \to R$ should be extended, on the other hand the inherent features of system evolution should be stabilized by removing the effect of short-term fluctuations for the system.

This and the regularization theory are consistent with prior information smoothing which often accepted for $f$.

Therefore, by retreating from the others demand and progressing the approximation for smoothed $f$, we are going to attain to the goal for forecasting of finance tine series.

There are three sorts of smoothing method that is, average smoothing method, exponent smoothing method, wavelet transformation smoothing method in the technical analysis of financial market

The smoothing data obtained from average smoothing method have certain time delay and the smoothing data obtained from exponent smoothing method make remove time delay.

These two sorts of method all are the typical smoothing method

Meanwhile, smoothing method by wavelet transformation is new technique developed in the last 20 years.

There is no the time delay in the reconstructed smoothing data as low-frequency spectrum by using wavelet transformation.

In the traditional time series analytical technique, it has been assumed that time series all is stationary in time.

But, actually time series data encountered with real world are the non stationary time series

Decomposing the time series over distinctive frequency channel by step after step in the processing method of time series data by wavelet, after that the decomposed data make have strong the stationary capacity.

Moreover, owing to wavelet decomposition make a smoothing action for time series data, it exhibit itself superiority in the processing for non-stationary random time series.

We have used the wavelet noise elimination procedure-wden in Wavelet Toolbox of MATLAB 7.8 and applied third order coiflet- coif3 with family of functions for wavelet.

By using time series obtained after which noise are removed to SVR time series model, we have progressed the forecast.

Then, in the comprehensive forecasting method, we have progressed the forecast by using fuzzy Markov model and weighted Markov model for the forecast error obtained from SVR and it made compensated to the forecast result of SVR.

In combinational forecast, we have used the combined weight determination method based on the least square combined weight determination method, the weight determination method by effective degree, the weight determination method by grey relation degree, the weight determination method by optimal grey relation degree.

Then, we made compared for the combinational forecast result.

## 3. The wavelet-least square support vector machine

How to obtain the algorithm which accuracy high and generalization ability strengthen is too the problem suggested very urgently.

In the paper [16], Support Vector Machine (SVM) is used and this is a sort of new algorithm.
This method have more advantage than another methods which performance criterion choose to minimize the empirical risk and it have optimal solution in the global range at the same time because of its algorithm is convex quadratic programming.
But, when data volume is large, traditional SVM algorithm demand great computational costs and there are also suggested some problems in the choice of kernel function.
In relation to this, least square support vector machine (LS-SVM) is a sort development of the traditional SVM.
There is that reason because of LS-SVM makes change to equality restriction for inequality restriction of SVM method and the error square and loss function make to empirical loss finally the problem make change to linear matrix problem.
This method is fast in the solving velocity than the algorithm convex quadratic optimization and less computational costs is required( [13,14]).
In this paper, we have applied this algorithm to construct wavelet kernel function.

### 3.1. LS-SVM prediction theory

Essentially, the forecast belongs to the regression problem.
That is, through the function estimation, the relational model between input value and out value is settled up and an ahead output value is forecasted by this model.
  LS-SVM algorithm is constructed as following( [2,3]).
Let $\{(x(1),y(1)),(x(2),y(2)),\cdots,(x(N),y(N))\}$ be an indicated training data.
Here $x(i) \in R^d$ are input arguments and $y(i) \in R$ are out arguments.
$y(i) = f(x(i))$ $(i=1,2,\cdots,N)$ and $f(x)$ ( $x = (x(1), x(2),\cdots, x(N))$ ) is an unknown function to be should estimate.
We make the nonlinear mapping $\phi: R^d \to H$ and function $f(x)$ to be should estimate choose as following:.

$$y = f(x) = w^T \phi(x) + b \tag{3}$$

Here $\phi$ is called the characteristic mapping and $H$ is called the characteristic space. $w$ is output vector in the space $H$ and $b \in R$ is bias
Then, LS-SVM method which estimates nonlinear function in the characteristic space becomes optimization problem as following.

$$\min_{w,b,e} J(w,e) = \frac{1}{2} w^T W + \frac{1}{2} \gamma \sum_{i=1}^{N} e(i)^2 \tag{4}$$

$$s.t. \quad y(i) = w^T \phi(x(i)) + b + e(i)$$

Here $e(i) \in R, i = 1, 2, \cdots, N$ is variable which express the error.

$J(w,e)$ is composed of regularization part $\frac{1}{2} w^T w$ and error square part $\frac{1}{2} \gamma \sum_{i=1}^{N} e(i)^2$.

Here $\gamma$ is positive constant.
Generally, solving the extreme value problem (4) is very difficult.

The reason is that $w$ can be infinite-dimensional.

Accordingly, we transform this problem to dual spaces and define Lagrange function such as;

$$L(w,b,e,\alpha) = J(w,e) - \sum_{i=1}^{N} \alpha_i [w^T \phi(x(i)) + b + e(i) - y(i)] \quad (5)$$

Here $\alpha$ is Lagrange's indeterminate multiplier

Then, the optimality condition is given as following:

$$\begin{cases} \dfrac{\partial L}{\partial w} = 0 \to w = \sum_{i=1}^{N} \alpha_i \phi(x(i)) \\ \dfrac{\partial L}{\partial b} = 0 \to \sum_{i=1}^{N} \alpha_i = 0 \\ \dfrac{\partial L}{\partial e(i)} = 0 \to \alpha_i = \gamma e(i) \\ \dfrac{\partial L}{\partial \alpha_i} = 0 \to w^T \phi(x(i)) + b + e(i) - y(i) = 0 (i=1,2,\cdots,N) \end{cases} \quad (6)$$

Eliminating $e(i)$ and $w$ and solving equation (6), following equation are obtained.

$$\begin{bmatrix} 0 & I_v^T \\ I_v & \Omega + \dfrac{1}{\gamma}I \end{bmatrix} \begin{bmatrix} b \\ \alpha \end{bmatrix} = \begin{bmatrix} 0 \\ Y \end{bmatrix}$$

Here

$$Y = [y(1), y(2), \cdots, y(N)] \quad , \quad I_v = [1,1,\cdots,1]^T \quad , \quad \alpha = [\alpha_1, \alpha_2, \cdots, \alpha_N] \quad ,$$

$$\Omega_{ij} = \phi(x(i))^T \phi(x(j)) = K(x(i), x(j)), i,j = 1,2,\cdots,N.$$

$K(x(i), x(j))$ is kernel function

This is called Mercer condition.

The estimating function of LS-SVM is given as following:

$$y(i) = \sum_{k=1}^{N} \alpha_k K(x(k), x(i)) + b \quad (7)$$

$\alpha_i$ and $b$ can be obtained from linear equation

There are various forms for kernel function but we take often as follows form([10])

(ⅰ) polynomial kernel function; $K(x,y) = (xy + c)^d$, c is constant and $d = 1,2,\cdots$

(ⅱ) RBF kernel function; $K(x,y) = \exp(\dfrac{-\|x-y\|^2}{2\sigma^2})$

(ⅲ) Sigmoid kernel function; $K(x,y) = \tanh[b(x \cdot y) - c]$

### 3.2. The construction of wavelet kernel function

In the papers [8,12], the input data make decomposed by using wavelet and the new decomposed coefficients are used with the input data. Then, combination of wavelet and SVM is realized by using RBF kernel function.

In this paper, we consider the method composing wavelet kernel function on the basic of certain frame

*[Essence of kernel function].*

Let $F$ be a frame and $R$ be a composition operator of $F$.

And let $R^*$ be a decomposition operator of the F.

If we put $Q = RR^*$, then $Q$ is bounded measurable frame( [11])

We assume that $\lambda$ is a positive operator and $G(x, y)$ is Green function of $\lambda^*\lambda$.

Then $G$ is the kernel of Mercer and
$$K(x, y) = <(\lambda K(x, \cdot), \lambda K(\cdot, y)>$$

$$(\lambda^*\lambda G(\cdot, y))(x) = \delta_y(x)$$

are satisfied.

In this formula $\delta_y(x)$ is estimation operator of $y$, $K(x, y)$ is kernel function to be should estimate.

Therefore, we can define kernel function such as;
$$K(\cdot, x): x \to \lambda G(\cdot, x)$$

Let $H^s$ be Sobolev space. We put $F = \{f_i\} \subset H = H^s(\Omega)$

Let $G(x, y) \in H^s(\Omega \times \Omega)$ be $Q$-Green function.

Making operator $\lambda = R^*$, we obtain $(QG(\cdot, y))(x) = \delta_y(x)$.

According to the character of $\delta_y(x)$, we obtain
$$G(x, y) = <G(\cdot, y), \delta_x(\cdot)>$$

Because of
$$G(x, y) = <G(\cdot, y), \delta_x(\cdot)> = <G(\cdot, y), QG(\cdot, x)> = <G(\cdot, y), RR^*G(\cdot, x)>$$

$$= <R^*G(\cdot, y), R^*G(\cdot, x)>_{l^2}$$

holds true.

Hence
$$K = G \tag{8}$$

Essentially, kernel function is Green function of decomposition and is composite normal operator under the some frame. This is very similar to wavelet decomposition and composition.

Therefore, again to think the solution of Green function ( kernel function ) under the wavelet frame is very easy.

*[Construction of kernel function by wavelet]*

The frame theory was proposed by Duffin etc at 1952 and it was researched based on concept of normal orthogonal basis

Its aim is to compose function $f$ from inner product of $f$ in Hilbert space with family of functions $\{\psi_k\}_{k \in K}$ ([7]).

We assume that family of functions $\{\psi_k\}_{k \in K}$ is given in Hilbert space $H$.

Then, if there exist $0 < A < B < \infty$ for all $f \in H$ such that

$$A\|f\|^2 \leq \sum_k |<f, \psi_k>|^2 \leq B\|f\|^2 \tag{9}$$

then family of functions $\{\psi_k\}_{k \in K}$ is called frame in the space $H$.

And $A$ and $B$ are called upper-limit and lower bound of frame respectively.

Especially, if $A = B$, then it is called strict frame and if $A = B = 1$, then it is called orthogonal frame.

If $\|\psi_k\| = 1$, then $\{\psi_k\}$ is normal orthogonal basis.

For arbitrary $f$ which belongs to $H$, we can decompose as following;

$$f = \sum_k <f, \psi_k> \psi_k = \sum_k <f, \psi_k> \bar{\psi}_k \tag{10}$$

Here $\bar{\psi}_k$ is conjugate frame of $\psi_k$.

Let $F = \{\psi_i\}$ be frame in the square integrable function space $L^2(\Omega)$ and $\{\lambda_i\}$ be increasing sequence of positive numbers. Then function $K(x, y)$ can be expressed such as

$$K(x, y) = \sum_i \lambda_i \psi_i(x) \psi_i(y) \tag{11}$$

The function defined by above formula is half positive value.

**[Theorem 3.1] (Mercer's Theorem)**

Let function $K(x, x')$ is symmetry in space $L^2$. Then the necessary and sufficient condition for which $K(x, x')$ is inner product in the characteristic space is to satisfy as following condition;

$$\iint_{R^d \times R^d} K(x, x') f(x) f(x') dx dx' \geq 0$$

for all $f$ which satisfies the condition $f \neq 0$, $\int_{R^d} f(\xi) d\xi < \infty$.

This theorem gives soon the process to judge and to construct the kernel function.
The above formula make to be used kernel function in SVM which it is to be satisfied the Mercer condition.

There is another kind of case.

If the parallel displacement invariant kernel function, for instance $K(x, x') = K(x - x')$ satisfies the Mercer's theorem, then it is admissible kernel function.

But actually to decompose the parallel displacement invariant kernel function to two functions is very difficult.

Following theorem gives the necessary and sufficient condition for which the parallel displacement invariant kernel function is kernel function in SVM.

**[Theorem 3.2]** The necessary and sufficient condition for which the parallel displacement invariant kernel function $K(x,x') = K(x-x')$ is admissible support vector kernel function is to satisfy Fourier transformation of $K(x)$ such as condition

$$F[K(w)] = (2\pi)^{\frac{d}{2}} \int_{R^d} e^{-i(w,x)} K(x)dx \geq 0$$

If theorem 3.1 and theorem 3.2 hold true, then the kernel function of SVM can be constructed.
Now, let us see the kernel function constructed by wavelet.
Let $\psi(x)$ be a generating function of wavelet.
Let $a$ be the flexible factor and $m$ be the parallel displacement factor in the generating function of wavelet $\psi(x)$.
If the wavelet frame is generated by this the generating function of wavelet, then the wavelet kernel function is constructed such as

$$K(x,x') = \prod_{i=1}^{d} \Psi[\frac{x(i) - m_i}{a_i}] \Psi[\frac{x'(i) - m'_i}{a'_i}] \qquad (12)$$

Here $x, x' \in R^d$; $x(i), x'(i), a_i, a'_i, m_i, m'_i \in R$, $a_i, a'_i \neq 0$.

*[Least-square wavelet Support Vector Machine (LS-WSVM)]*
In the above, we have made the kernel function of general SVM based on wavelet.
If the wavelet kernel function is determined, then the model is soon determined.
In this paper, to construct the parallel displacement invariant kernel function, we have choused Mexican Hat wavelet kernel function $\psi(x) = (1-x^2)\exp(-\frac{x^2}{2})$.

By above mentioned argument, the parallel displacement invariant wavelet kernel function composed by Mexican Hat wavelet is given such as;

$$K(x,x') = K(x-x') = \prod_{i=1}^{d} \Psi[\frac{x(i) - x''(i)}{a_i}] = \prod_{i=1}^{d} [1 - \frac{\|x(i) - x'(i)\|^2}{a_i^2}] \exp[-\frac{\|x(i) - x'(i)\|^2}{a_i^2}] \qquad (13)$$

Here $x(i), x'(i), a_i \in R$, $a'_i \neq 0$, $x, x' \in R^d$.
This is a sort of admissible kernel function of SVM.

Now, we assume that $G = \{x(i), y(i)\}_{i=1}^{N}$ are training samples and $x_i \in R^d$ are input vectors.

Here $N$ is the whole number of data and $y(i) \in R$ is the expected value.
We assume that the numbers of support vector included into the input vector is $M$ respectively. And we denote those by $SV_1, SV_2, \cdots, SV_M$.
Then, by means of wavelet transformation, corresponding wavelet frame $\Psi(x(1))$, $\Psi(x(2))$, $\cdots, \Psi(x(M))$ satisfying Mercer condition is obtained.
Therefore, we can obtain the kernel function corresponding to it given by the formula (13).
Accordingly, we can obtain the LS-WSVM approximate function which appropriate to the regression problem such as

$$y = \sum_{k=1}^{N} \alpha_k \prod_{i=1}^{d} [1 - \frac{\|x_k(i) - x'_k(i)\|^2}{a_k'^2}] \exp[-\frac{\|x_k(i) - x'_k(i)\|^2}{2a_k'^2}] + b \qquad (14)$$

In this formula $x_k(i)$ is $i$th component in $k$ number of training data.

So, we have constructed a new regression method by using wavelet kernel function and LS-SVM

That is, we have constructed the Least-square Support Vector Machine (LS-WSVM) based on wavelet. Essentially, LS-WSVM is a sort of form for LS-SVM.

To choice $N \times d$ number parameters $a'_k$, $i = 1,2,\cdots,d; k = 1,2,\cdots,N$ is difficult because of the parameters of kernel for LS-SVM can't be optimized.

Therefore, to decrease and simplify the numbers of parameters, we put $a'_k = a$.

Then, the numbers of parameters for the kernel function become one.

The super parameter $a$ of wavelet kernel function can be choosed by using the cross-validity method. But, in this paper we have used the genetic algorithm.

The support vector machine SVM is strong nonlinear system. The less change of its intrinsic parameter exerts on the effect in the adapted performance of the system.

Generally, the parameter $\sigma$ of the kernel function for SVM has been established in initial moment next it keeps on constancy.

But, owing to the financial market system is strange and complicate besides there is the time-varying law , clearly it don't conform with to use the invariant kernel parameter at direct series forecast actually.

Therefore, in this paper we have proposed the method adjusting the kernel function parameters of SVM automatically by using the genetic algorithm.

The genetic algorithm is a kind of global optimal search algorithm simulating the process of organic evolution.

Their objective functions not only don't demand the continuity but also don't demand the differentiability, have only to demand calculability of the problem.

Besides, its exploration has been progressed in the integrated solution space from the beginning to the finishing.

Therefore, the global optimal solution has been obtained easily.

Suppose that the test length is $T$ and the sample length is $N$.

Suppose $\hat{y}_1(i)$ $(i = 1,2,\cdots,T)$ is the predicted value for the $T$ number of test data and $y_1(i)$ $(i = 1, 2,\cdots,T)$ is the $T$ number of actual value for the test data.

And we introduce the notation $\hat{y}_1 = (\hat{y}_1(1), \hat{y}_1(2),\cdots, \hat{y}_1(T))$, $y_1 = (y_1(1), y_1(2),\cdots, y_1(T)))$

$$RMSE_1 = \left\| \frac{\hat{y}_1 - y_1}{T} \right\|_{L^2} = \sqrt{\frac{\sum_{i=1}^{T}[\hat{y}_1(i) - y_1(i)]^2}{T}}$$

Then, we define the calibration error such as;

$$RMSE_1 = \left\| \frac{\hat{y}_1 - y_1}{T} \right\|_{L^2} = \sqrt{\frac{\sum_{i=1}^{T}[\hat{y}_1(i) - y_1(i)]^2}{T}}$$

Suppose that $\hat{y}_1(i)$ $(i=1,2,\cdots,N)$ is the predicted value for the $N$ number of fitting data and $y_1(i)$ $(i=1, 2,\cdots,N)$ is the $N$ number of actual value for the fitting data.
And we introduce the notation $\hat{y}_2 = (\hat{y}_2(1), \hat{y}_2(2),\cdots, \hat{y}_2(N))$, $y_2 = (y_2(1), y_2(2),\cdots, y_2(T))$
Then, we define the error of fitting such as;

$$RMSE_2 = \left\| \frac{\hat{y}_2 - y_2}{N} \right\|_{L^2} = \sqrt{\frac{\sum_{i=1}^{N}[\hat{y}_2(i) - y_2(i)]^2}{N}}$$

$$fitness = -(\eta_1 RMSE_1 + \eta_2 RMSE)$$

With due regard to the calibration error and the error of fitting comprehensively, we have defined the fitness function such as

$$RMSE_2 = \left\| \frac{\hat{y}_2 - y_2}{N} \right\|_{L^2} = \sqrt{\frac{\sum_{i=1}^{N}[\hat{y}_2(i) - y_2(i)]^2}{N}}$$

Here $\eta_1$ and $\eta_2$ denote the weight coefficients. We put $\eta_1 = 0.3, \eta_2 = 0.7$.
Let note that the fitness function is the inverse of the predicted error.
Therefore, to maximize the fitness function corresponds to minimize the predicted error.
When the genetic algorithm arrives at the stop standard, an individual with the greatest fitness degree is corresponded to optimal parameter of SVM.

## 4. Forecasting by fuzzy Markov-support vector regression and weight Markov

We assume that $\hat{X}_t$ is the fitting curve which is obtained through forecasting for the financial time series $Y_t$ by using wavelet-support vector regression.
In order to progress Markov prediction, it is need to make several state by dividing the relative error of given time series.
In due consideration of the actual meaning of the time series, in this paper we have generated a new time series $Z_t = (Y_t - \hat{X}_t)/Y_{t-1}$ $(t = 2,3,\cdots,N)$ which the fitting curve $\hat{X}_t$ make reference
Here $Y_{t-1}$ is one step of time delay value.
In accordance with the distribution of the random sequence $Z_t$, it is divided by $k$ pieces of state and is taken its partition value. That is,
$$e_1 = [m_0, m_1], e_2 = [m_1, m_2],\cdots, e_{k-1} = [m_{k-2}, m_{k-1}], e_k = [m_{k-1}, m_k]$$
Here $m_i \geq m_{i-1}$.
Then, the sequence $Z_t$ is the random sequence which $E = \{e_i\}$ $(i=1,2,\cdots,k)$ become its state space.
Markov process is a field of random process and its greatest character is the "un-posterior effectiveness".
That is, the "future" and the "past" of the given some random process is independent under the condition of "present".
Markov chain is the Markov process which its state and time parameter all are discrete and its mathematical notation is given as follows.

We assume that sequence $\{X(t), t \in T\}$, $T = \{0,1,2,\cdots\}$ of probability space $(\Omega, F, P)$ was defined
The state space $I = \{0,1,2,\cdots\}$ is called Markov chain, if it satisfied the condition such as; that is, for arbitrary positive integers $l, m, k$ and nonnegative integer $j_l > \cdots > j_2 > j_1 (m > j_l)$, $i_{m+k}$, $i_m$, $i_{jl}$, $\cdots$, $i_{j2}$, $i_{j1}$

$$P\{X(m+k) = i_{m+k} | X(m) = i_m, X(j_l) = i_m X(m) = i_{jl}, \cdots, X(j_2) = i_{j2}, X(j_1) = i_{j1}\} =$$
$$= P\{X(m+k) = i_{m+k} | X(m) = i_m\} \tag{15}$$

holds true.
Here the left-hand side of expression (15) should have the meaning, that is
$$P\{X(m) = i_m, X(j_l) = i_m X(m) = i_{jl}, \cdots, X(j_2) = i_{j2}, X(j_1) = i_{j1}\} > 0$$
There are many definitions for the properties and characters of Markov chain.
We don't mention for that one by one.

In the practical application, we consider the homogeneous Markov chain, that is, for arbitrary $k, n \in N^+$

$$P_{ij}(n, k) = P_{ij}(k), \quad i, j = 0,1,2,\cdots, \tag{16}$$

holds true.

Here $P_{ij}(n, k)$ denote the probability which $n$-step state equals to $i$ and the transition probability transited to the state $j$ by passed $k$-step is given by it.

The transition probability transited form $i$ to $j$ by pass $k$-step is denoted by $P_{ij}(k)$. That is,

$$p_{ij} = p\{x_{k+1}^{(0)} = j | x_k^{(0)} = i\} = n_{ij} / n_i = M_{ij} \quad (i, j = 1, 2, 3, \cdots, m; k = 1, 2, \cdots, (N-1)).$$

The homogeneous Markov chain $X(t)$ is perfectly determined by its initial distribution $\{P(i), i = 0,1,2,\cdots\}$ and its state transition probability matrix (that is, it is composed by the state transition probability $P_{ij}(i, j = 0,1,2,\cdots)$).

 *[Testing of Markov property]*
We assume that $n_i$ is the number which down at state $e_i$ among the $N$ number points of given random sequence $Z_t$. We denote the frequency $n_{ij}$ of transition from $e_i$ to $e_j$ for the state $Z_1, Z_2, \cdots, Z_N$ by passed one step.
Then, we accept the statistical quantity such as;

$$P_{ij} = \frac{n_{ij}}{n_i}, \quad P_{0j} = \frac{\sum_{i=1}^{k} n_{ij}}{\sum_{i=1}^{k} n_i}, \quad \chi^2 = 2 \sum_{i=1}^{k} \sum_{j=1}^{k} n_{ij} \left| \log \frac{P_{ij}}{P_{0j}} \right|$$

Then, when $N$ is comparatively large, the statistical quantity $\chi^2$ is according to $\chi^2$-distribution with the degree of freedom $(k-1)^2$.([3])

When giving the degree of confidence $\alpha$, if $\chi^2 > \chi_\alpha^2(m-1)^2$, then we confirm that the random sequence $Z_t$ have Markov's property, unless, the sequence haven't Markov's property.

*[Fuzzy Markov forecasting model]*

Suppose $U$ is the numerical range which random variable of Markov chain adopt and we construct the fuzzy state set $S_1, S_2, \cdots, S_l$

If for arbitrary $u \in U$ the condition

$$\sum_{m=}^{l} \mu_{S_m}(u) = 1$$

is satisfied, and then $\mu_{S_m}(u)$ is called the member degree of the fuzzy state $S_M$ for numerical value $U$.

The value $Z_t$ which the fuzzy variable adopts in some time certainly didn't revert to some state by all means, but it is split by another member degree and belongs to each fuzzy state.

That is $Z_t$ is split and belongs to each fuzzy state.

**[Definition 4.1]** Suppose $\mu_{S_i}(Z_t) \cdot \mu_{S_j}(Z_{t+1})$ is the fuzzy state transition coefficient from the state $S_i$ to state $S_j$ when time is turned from $t$ into $t+1$.

Then,

$$a_{ij} = \sum_{t=1}^{N-1} \mu_{S_i}(Z_t) \cdot \mu_{S_j}(Z_{t+1})$$

is called fuzzy transition frequency number from the state $S_i$ to the state $S_j$.

When the state $Z_t$ belongs to $S_i$ with degree of member $\mu_{S_i}(Z_t)$ and belongs to $S_j$ with degree of member $\mu_{S_j}(Z_t)$, the transition order is only expressed by the product between degrees of member $\mu_{S_i}(Z_t) \cdot \mu_{S_j}(Z_{t+1})$.

Owing to the fuzzy transition probability from the state $S_i$ to the state $S_j$ is denoted by

$$P_{ij} = \frac{a_{ij}}{\sum_{j=1}^{} a_{ij}}, ((i, j = 1, 2, \cdots, M)$$

, then the time series time series which we are going to establish is given such as;

$$\hat{Y}_t = \hat{X}_t + \sum_{i=1}^{k} \mu_{S_i}(Z_{t-1}) \sum_{j=1}^{k} \frac{1}{2}(m_{i-1} + m_i) P_{ij} Y_{t-1}, \quad t = 1, 2, \cdots, N$$

Here, $\hat{X}_t$ is the predicted value obtained by using wavelet –support vector.

*[Forecasting model of weighted Markov chain]*

The autocorrelation coefficient of each order describes the correlation and its strength and weakness between the finance time series value in the every delay time, because of the finance time series is the random variable quantity depending mutually.

Therefore, in forecasting of finance time series, we can consider as following method.

The first, we choice the finance time series of several time stage before present time stage and can progress the forecast of finance time series of present time stage by using some method.

Then, we find out the weight added sum by the strong and the weak of correlation between before each time stage and present time stage.

That is, we can arrive at the forecast goal by using the information sufficiently and rationally.

This is immediately the basic idea of the forecasting of weighted Markov chain.

The basic method and procedure of the forecasting of weighted Markov chain is established such as:

Step 1. For the financial time series, calculate the autocorrelation coefficient of each order $r_k$, that is,

$$r_k = \sum_{t=1}^{n-k}(x_t - \bar{x})(x_{t+k} - \bar{x}) / \sum_{t=1}^{n}(x_t - \bar{x})^2 \qquad (17)$$

Here, $r_k$ denote the autocorrelation coefficient of $k$ time stage (delay time $k$) $x_k$ denotes the financial time series value of $t$ time stage. And $\bar{x}$ denotes the average value of financial time series $n$ denote the length of financial time series.

Step 2. progress the normalization for the autocorrelation coefficient of each order.

$$w_k = |r_k| / \sum_{k=1}^{m}|r_k| \qquad (18)$$

These make $m$ of Markov chain for every kind delay time (step length).

Here $m$ is the optimal order calculated by forecast demand.

Step 3. Determine the grading standard of the financial time series (this is corresponded to the determination of state space for Markov chain).

This is progressed based on the long and short of data sequence as well as the demand of problem. For example, we can divide grading standard by the class of $i_l$ number.

This is corresponded to the state space $I = \{i_1, i_1, \cdots i_l\}$.

Step 4. On the ground of the grading standard established by procedure 3), the state of grading standard for each time stage is determined.

Step 5. By progressing the statistical calculate for the results obtained by procedure 4), we can obtain transition probability for Markov chain of each other differential steps.

This determines the probability law of the state transition process for the financial time series

Step 6. If each financial time series before some time stage is selected as an initial state and is coupled the state transition probability matrix corresponded to it, then the state probability $P_i^{(k)}$ at this time stage for the financial time series is immediately forecasted.

Here, $i, i \in I$ is the state and $k$ ($k = 1, 2, 3, \cdots$) is the delay time (step).

Step 7. Making the weight added sum for each forecast probability of same state, it become the forecast probability to be faced in this state for the financial time series, that is

$$P_i = \sum_{k=1}^{m} w_k P_i^{(k)} \quad (19)$$

Thus, finally $i$ corresponded to $\max\{P_i, i \in I\}$ become immediately the forecast state of this time stage for the financial time series.

After that obtained the financial time series value occurred at this time stage, it inserts in the original time series and repeat the procedure 1)-7), then the forecast value in next time stage for financial time series can be obtained.

Therefore, we make the forecasting model by the fuzzy Markov-support vector regression and the weighted Markov such as;

$$\hat{Y}_t = \frac{1}{2}(\hat{Y}_t^{(1)} + \hat{Y}_t^{(2)})$$

Here, $\hat{Y}_t$ is the comprehensive forecast value and $\hat{Y}_t^{(1)}$ are the forecast value by the fuzzy Markov-support vector regression forecasting model and $\hat{Y}_t^{(2)}$ the forecast value by the weighted Markov forecasting model.

The forecasting model obtained by above methods is called wavelet-support vector regression-Markov (Wavelet-SVR-Markov) forecasting model.

## 5. The comprehensive forecasting model

Practically, there are various sorts of models which can be used to forecast generally.

When encounter to the problem how choice the actual forecast value from different forecast values obtained by using various types of "possible" models, the problem such that predictor use what kind of model and choose what sort of predicted value is importantly suggested.

In fact, each model reflects only some aspect of the information under this circumstance.

To neglect the model which can be used in some degree latently means that some wrathful information can be neglect. And if any single model is only to choose, then global character of time series evolution can't be described perfectly.

In order to avoid these problems, to making the comprehensive forecast is suitable for predictor by considering the combined forecast method.

The idea of comprehensive forecast is such that the predicted values obtained from various kinds of models how make to combine with organized method by the added accuracy.

If each forecast method is well combined, then not only a matter for regret such that the single individual model throw away some aspect of useful information can be avoided but also the randomness of forecast can be weakened and the accuracy can be enhanced.

At the present, the idea of comprehensive forecast is widely used in various kinds of forecasts practically.

If the forecast is only made by traditional single model, then there can be some shortcoming.

For instance, the information source can't be extensive and there can be the sensitivity for the model setting form etc.

After suggesting first the method of comprehensive forecast in paper [1], it not only receive importance from many predict experts ,but also become heat point task in the forecast field([1-11]).

The comprehensive forecast is immediately the method presenting the effective forecast result by combining the individual forecast result obtained from single forecast model with the suitable weighted average form.

The most key problem in the comprehensive forecast is how to find good weight coefficient and to enhance more effectively for the comprehensive forecast model.

At present, the most many comprehensive forecast methods suggested in actual application and theoretical research is to calculate the weight coefficient of the comprehensive forecast methods by using a kind of absolute error make as optimal criteria.

Therefore, the researches of comprehensive method don't be completed yet and an optimal comprehensive forecast model built various kinds of combined criteria is demanded.

So, to enrich the comprehensive forecast theory and methods is demanded in going one more step forward.

In paper [6] an optimal comprehensive forecast model which the error sum of squares arrive at minimum under the condition of unless non-negative constraint is built and the condition has been judged which a simple average method is whether the recessive comprehensive forecast or dominant comprehensive forecast by using the property of information matrix for the comprehensive forecast absolute error.

In paper [7], going one more step forward. an optimal comprehensive forecast model of the error sum of squares under the condition of non-negative constraint condition has been studied.

In paper [8], a comprehensive forecast model based on the forecast effective degree has been studied and the solving method of linear programming for it's has been given.

In paper [9], the properties of the comprehensive forecast model based on the forecast effective degree have been studied. In paper [10], another one of new pathway for the comprehensive forecast method has been presented.

That is, an optimal comprehensive forecast model based on dependence indicator has been presented. These models include the comprehensive forecast model maximizing grey relation degree, the comprehensive forecast model maximizing correlation coefficient, the comprehensive forecast model maximizing narrow-angle cosine etc.

These methods don't directly consider the magnitude of prediction error. Therefore, these methods have comparatively the great disparity over the traditional comprehensive forecast model

But, in paper [9] it has been explained that the comprehensive forecast method based on the correlation in only a standpoint of an actual proof analysis can be obtained a good comprehensive forecast effect relatively.

In this paper, the effectiveness of the comprehensive forecast method based on the grey relation degree upon comparison with the comprehensive forecast model based on the maximization for grey relation degree suggested in paper [10] has been explanted mainly.

*[Dominant forms of the typical comprehensive forecast]*

By considering the principle of comprehensive forecast, there are three kind of mutual another comprehensive forecast method that is, arithmetic average combination method, geometric mean combination method and harmonic average combination method.

Those combination formulas are given such as respectively

$$\hat{y}(t) = k_1 \hat{y}_1(t) + k_2 \hat{y}_2(t), \quad (t = 1,2,\cdots,N)$$

$$\hat{y}(t) = \hat{y}_1^{k_1}(t) + \hat{y}_1^{k_2}(t), \quad (t = 1,2,\cdots,N)$$

$$\hat{y}(t) = \frac{1}{k_1/\hat{y}_1(t) + k_2/\hat{y}_2(t)}, \quad (t = 1,2,\cdots,N)$$

Here $N$ is total number of forecast data which we are going to obtain $\hat{y}_1(t), \hat{y}_2(t)$ are the predicted values by using the model 1 and model 2 respectively. $k_1, k_2$ are the weight coefficients of two forecast models respectively.

The most important problem in the comprehensive forecast is to find the weight coefficients $k_1, k_2$ by how method. The most common used method in the comprehensive forecast is the arithmetic average combination forecast form, that is,

$$\hat{y}(t) = \sum_{i=1}^{m} k_i \hat{y}_i(t), \quad (t = 1,2,\cdots,N)$$

Here $\hat{y}_i(t)$ $(i = 1,2,\cdots,m)$ is the predicted value by the forecast method of $i$ th model and $m$ is number of forecast model. $k_i$ is the weight coefficient of $i$ th model.

*[Some weight determination methods of comprehensive forecast model]*

In the determination methods of combinatorial weight for the traditional comprehensive forecast model there are the determination methods of combinatorial weight based on averaging, the determination methods of combinatorial weight based on standard deviation, and based on deviation coefficient etc. Concretely, those methods are given such as;

（ⅰ）The determination methods of combinatorial weight based on averaging.

In this method, the weight $w_j$ is taken such as;

$$w_j = \frac{1}{m} \quad (j = 1,2,\cdots,m) \tag{20}$$

Owing to take the average in this method, all models entering in main combination have been dealt with uniformity and equality

This method has been often used when the predictor don't select a key which some model have an advantage than another model in the models taking part in the comprehensive forecast.

（ⅱ）The determination methods of combinatorial weight based on standard deviation

In this method, the weight $w_j$ is taken such as;

$$w_j = \frac{S - S_j}{S} \cdot \frac{1}{m-1}, \quad S = \sum_{j=1}^{m} S_j \quad (j = 1,2,\cdots,m) \tag{21}$$

Here $S_j$ is the standard deviation of $j$'th model.

If this model has been used, then the model which have minimum standard deviation should been taken the greatest weight in the comprehensive forecast model.

(ⅲ) The determination methods of combinatorial weight based on deviation coefficient.

We assume that there are $n$ numbers forecast points.

Let $\hat{y}_j^i$ be the predicted value in the point $i(i=1,2,\cdots,n)$ by model $j(j=1,2,\cdots,m)$ and $\bar{y}^i$ be the average of $m$ numbers of predicted value in the point $i$.

Then, the weight $w_j$ is taken such as ;

$$w_j = \frac{d-d_j}{d} \cdot \frac{1}{m-1} \quad (j=1,2,\cdots,m) \tag{22}$$

Here $d_j$ is deviation coefficient and $d$ is sum of $d_j$ defined as below

$$d_j = \frac{1}{n}\sqrt{\sum_{j=1}^{n}(\hat{y}_j^i - \bar{y}^i)^2} \quad (j=1,2,\cdots,m), \quad d = \sum_{j=1}^{m} d_j .$$

Now, let us consider the determination methods of combinatorial weight based on least square method, the determination methods of combinatorial weight based on effective degree, the determination methods of combinatorial weight based on grey relation degree, the determination methods of combinatorial weight based on optimal grey relation degree, the determination methods of combinatorial weight based on the theory of rough set etc. These methods have been made attention recently.

**5.1. The determination methods of combinatorial weight based on least square method**

Let $y = \{y(1), y(2),\cdots, y(N)\}$ be original data series and $\hat{y}_j(t)$ $(j=1,2,\cdots;m)$ be predicted value of the $j$'th method in $t$ time, let $w_j$ $(j=1,2,\cdots;m)$ be its weights.

Here suppose $w_j$ satisfies as following condition;

$$\sum_{j=1}^{m} w_j = 1, \quad w_j \geq 0 \ (j=1,2,\cdots;m) \tag{23}$$

Now, we construct the comprehensive forecast model such as;

$$\hat{y}(t) = \sum_{j=1}^{m} w_j \hat{y}_j(t) \tag{24)}$$

In order to determine the weight $w_j$, the optimization criterion has been choused with the error square sum, that is

$$\min J = \sum_{t=1}^{N}[\hat{y}(t)-y(y)] \qquad (25)$$

Then, the comprehensive forecast model become above optimization problem.

Now, we denote the error such as;

$$\varepsilon_{jt} = \hat{y}_j(t) - y_j(t), (j=1,2,\cdots,m; t=1,2,\cdots,N) \qquad (26)$$

The error square sum is given by

$$h_{jt} = \sum_{t=1}^{N}\varepsilon_{jt}^2, \quad (j=1,2,\cdots,m) \qquad (27)$$

Now, let us $\varepsilon_{it}$ and $\varepsilon_{jt}$ are the prediction errors of $i$'th method and $j$'th method respectively

Then, $\varepsilon_{it}$ and $\varepsilon_{jt}$ is mutually independent respectively

Therefore, we have;

$$h_{ij} = \sum_{t=1}^{n}\varepsilon_{it}\cdot\varepsilon_{jt} \approx 0, \quad (i,j=1,2,\cdots,m) \qquad (28)$$

Then, we introduce the notation such as

$$H = diag[h_{ij}], \quad W = [w_1, w_2, \cdots, w_m]^T, \quad E = [1,1,\cdots,1]^T \qquad (29)$$

From above expression, the matrix form the comprehensive forecast optimization model has been obtained such as;

$$\min J = W^T H W$$
$$s.t. \begin{cases} E^T W = 1 \\ W \geq 0 \end{cases} \qquad (30)$$

Let us accept Lagrange multiplier $\lambda$ and construct Lagrange function such as

$$L = W^T H W + \lambda(E^T W - 1) \qquad (31)$$

According to Kuhn-Tucker condition, we have

$$\begin{cases} \dfrac{\partial L}{\partial W^T} = 2HW + \lambda E = 0 \\ \dfrac{\partial L}{\partial \lambda} = E^T W - 1 = 0 \end{cases} \qquad (32)$$

Solving these equations, we obtain

$$w_j = \dfrac{1}{h_{jj}\sum_{i=1}^{m}\dfrac{1}{h_{ii}}}, \quad \lambda = -\dfrac{2}{\sum_{i=1}^{m}\dfrac{1}{h_{ii}}}$$

## 5.2. The determination methods of combinatorial weight based on the effective degree

Let us introduce the conception of effective degree of the forecast technique

These indicators have certain reasonableness because the forecast accuracy is reflecting the effectiveness of the forecast technique.

To explain a discussion briefly, let us $\hat{y}_1(t)$ and $\hat{y}_2(t)$ $(t=1,2,\cdots N)$ are the predicted value obtained by two kind of forecast method respectively.

Then, an idea of the determination methods of combinatorial weight by the effective degree has been given such as;

Now, if we introduce

$$A(t) = 1 - \left|\frac{y(t) - \hat{y}(t)}{y(t)}\right| = 1 - \left|\frac{y(t) - k_1\hat{y}_1(t) - k_2\hat{y}_2(t)}{y(t)}\right|$$

then $A(t)$ constitutes the accuracy of the comprehensive forecast

Let us define the average value $E$ and average square deviation $\sigma$ of $A(t)$ such as respectively

$$E = \frac{1}{N}\sum_{t=1}^{N} A(t), \quad \sigma = \frac{1}{N}(\sum_{t=1}^{N}(A(t)-E)^2)^{\frac{1}{2}}$$

Then, the effective degree of the comprehensive forecast method has been defined such as

$$S = E(1-\sigma)$$

It show that the greater $S$ is, the more higher the accuracy of the forecast model, the more stable and the more and more effective the model

There is comparatively the complicated method which $k_1$ and $k_2$ find out by optimizing of $S$, but there is also simple method which they find out by using physical meaning.

Now, assume that $A_1(t)$ and $A_2(t)$ are the forecast accuracy series using model 1 and model 2 respectively

From above expression, we can find the effective degree $S_1$ and $S_2$ of model 1 and model 2 respectively

Then, by standardizing the $S_1$ and $S_2$, we put the weight coefficient $k_1$ and $k_2$ such as

$$k_j = \frac{S_j}{\sum_{i=1}^{2} S_i}, \quad j=1,2$$

**5.3. The determination methods of combinatorial weight based on grey relation degree**

Before considering the determination methods of combinatorial weight based on grey relation degree, let us consider the concept of the degree of grey incidences

In systems analysis, after the characteristic quantities which describe systems' behaviors well have been chosen, we need to clarify all the factors that effectively affect the systems' behaviors.

**[Definition 5. 1]**. Assume that $X_i$ is a systems' factor with the k'th observation value being $x_i(k)$ ($k=1,2,\cdots, N$). Then $X_i = (x_i(1), x_i(2),\cdots, x_i(N))$ is called a behavioral sequence of the factor $X_i$

**[Definition 5.2]**. Assume that $X_0 = (x_0(1), x_0(2),\cdots, x_0(N))$ is a sequence of data representing a system's characteristics, and $X_i = (x_i(1), x_i(2),\cdots, x_i(N))$ ($i=1,2,\cdots,m$) are sequences of relevant factors. $X = \{X_i | i=1,2,\cdots,m\}$ is called the set of elements of relevant factors.

For a given real number $\gamma(x_0(k), x_i(k))$ if we introduce expression

$$\gamma(X_0, X_i) = \frac{1}{n}\sum_{k=1}^{n}\gamma(x_0(k), x_i(k))$$

then it satisfies the axioms 4 for grey relation ([41]).

Then $\gamma(X_0, X_i)$ is called the grey relation degree of $X_i$ with respect to $X_0$, and $\gamma(x_0(k), x_i(k))$ the relation coefficient of $X_i$ with respect to $X_0$ at point $k$.

Assume that $X_0 = (x_0(1), x_0(2), \cdots, x_0(N))$ is a sequence of data representing a system's characteristics, and $X_i = (x_i(1), x_i(2), \cdots, x_i(N))$ $(i = 1, 2, \cdots, m)$ are sequences of relevant factors.

For $\rho \in (0,1)$, define

$$\gamma(x_0(k), x_i(k)) = \frac{\min\limits_{i}\min\limits_{k}|x_0(k) - x_i(k)| + \rho \max\limits_{i}\max\limits_{k}|x_0(k) - x_i(k)|}{|x_0(k) - x_i(k)| + \rho \max\limits_{i}\max\limits_{k}|x_0(k) - x_i(k)|}$$

and

$$\gamma(X_0, X_i) = \frac{1}{n}\sum_{k=1}^{n}\gamma(x_0(k), x_i(k))$$

Then, $\gamma(X_0, X_i)$ satisfies the four axioms for grey incidences, where $\rho \in (0,1)$ is called the identification coefficient. Generally we take $\rho = 0.5$

The grey relation degrees are numerical characteristics for the relationship of closeness between two sequences.

When analyzing systems, and studying relationships between systems' characteristic behaviors and relevant factors' behaviors, we are mainly interested in the ordering of the degrees of incidence between the systems' characteristic behaviors and each relevant factor's behavioral sequence.

But, the numerical magnitude of the grey relation degree is incomplete.

The order of grey relation is complete ordering relation over $X$

In the research of time series, $X_0 = (x_0(1), x_0(2), \cdots, x_0(N))$ is called a sequence of system's characteristics and $X_i = (x_i(1), x_i(2), \cdots, x_i(N))$ $(i = 1, 2, \cdots, m)$ is called a sequences of comparison

Let us consider the determination methods of combinatorial weight based on grey relation degrees

We assume that an actual value of time series $\{y(k) | k = 1, 2, \cdots, N\}$ is given.

Let us there are $m$ kind of forecast method.

Let $\hat{y}_i(k)$ $(i = 1, 2, \cdots, m, k = 1, 2, \cdots, N)$ be predicted value of the $i$'th method in $k$ time.

If we put $e_i(k) = y(k) - \hat{y}_i(k)$, then $e_i(k)$ is the predicted error of $k$ time for $i$'th forecast method.

Now, we put comparison sequence $X_i \equiv E_i = (e_i(1), e_i(2), \cdots, e_i(N))$ and reference sequence $X_0 = (x_0(1), x_0(2), \cdots, x_0(N)) = (0, 0, \cdots, 0)$ and find the grey relation degree $\gamma(X_0, X_i)$ $(i = 1, 2, \cdots, m)$. Then we obtain the vector of grey relation degree $\Gamma = (\gamma(X_0, X_1), \gamma(X_0, X_2), \cdots, \gamma(X_0, X_m))$

After normalizing the vector of the grey relation degree $\Gamma$, we put it is combinatorial weight vector $W = (w_1, w_2, \cdots, w_m)$. The greater weight $(i = 1, 2, \cdots, m)$, then the effecter for forecast.

## 5.4. The determination methods of combinational weight based on optimal grey relation degree

We assume that an actual value of time series $\{y(k) | k = 1, 2, \cdots, N\}$ is given.

We assume that there are $m$ kind of forecast method.

Let $\hat{y}_i(k)$ $(i = 1, 2, \cdots, m, k = 1, 2, \cdots, N)$ be predicted value of the $j$'th forecast method in $k$ time $(j = 1, 2, \cdots, m, t = 1, 2, \cdots, N)$.

We introduce some conceptions such as ([12]);

**[Definition 5. 3]** We put

$$\gamma_{0j} = \frac{1}{N} \sum_{t=1}^{N} \frac{\min_{1\leq j \leq m} \min_{1\leq t \leq N} |e_j(t)| + \rho \max_{1\leq j \leq m} \max_{1\leq t \leq N} |e_j(t)|}{|e_j(t)| + \rho \max_{1\leq j \leq m} \max_{1\leq t \leq N} |e_j(t)|}$$

Then, $\gamma_{0j}$ is called the grey relation degree of the predicted value sequence

$\{\hat{y}_j(t) | t = 1, 2, \cdots, N\}$ by $j$'th forecast method between the actual value sequence

$\{y(t) | t = 1, 2, \cdots, N\}$. Here $\rho \in (0,1)$ is called the identification coefficient. Generally we take $\rho = 0.5$.

$e_j(t) = y(t) - \hat{y}_j(t)$ is the predicted error of $k$ time for $i$'th forecast method. Obviously, we obtain

$0 \leq \gamma_{0j} \leq 1$.

On the ground of definition 3, the d grey relation degree the predicted value sequence by a certain kind of forecast method between the actual value sequence is equal 1 if only and if the predicted value sequence and the actual value sequence is perfectly equal.

Let

$$\hat{y}(t) = w_1 \hat{y}_1(t) + w_2 \hat{y}_2(t) + \cdots w_m \hat{y}_m(t), \quad t = 1, 2, \cdots, N$$

be the predicted value of $y(t)$ by the comprehensive forecast method.

Here $w_1$, $w_2$, $\cdots$, $w_m$ are the weight coefficient of $m$ kind of forecast method and

$$\sum_{j=1}^{m} w_j = 1, w_j \geq 0 \quad (j = 1, 2, \cdots, m)$$

From above expression, we find the inequality as;

$$\min_{1\leq j \leq m} y_j(t) \leq \hat{y}(t) \leq \max_{1\leq j \leq m} y_j(t), t = 1, 2, \cdots, N$$

Let $e_t$ be the predicted error of $k$ time for $i$'th kind of forecast method.
Then, we obtain

$$e(t) = y(t) - \hat{y}(t) = y(t) - \sum_{j=1}^{m} w_j \hat{y}_j(t) = \sum_{j=1}^{m} w_j (y(t) - \hat{y}_j(t)) = \sum_{j=1}^{m} w_j e_j(t), t = 1, 2, \cdots, N$$

Let $\gamma$ be the grey relation degree of the comprehensive forecast method.
From the definition 5.3 and above expressions, we obtain

$$\gamma = \frac{1}{N} \sum_{t=1}^{N} \frac{\min_{1\leq j \leq m} \min_{1\leq t \leq N} |e_j(t)| + \rho \max_{1\leq j \leq m} \max_{1\leq t \leq N} |e_j(t)|}{\left|\sum_{j=1}^{m} w_j e_j(t)\right| + \rho \max_{1\leq j \leq m} \max_{1\leq t \leq N} |e_{jt}(t)|}$$

Obviously, the grey relation degree $\gamma$ is function of weight coefficient vector
$W = \{w_1, w_2, \cdots, w_m\}^T$ for the various kinds of forecast method.
Therefore, we denote $\gamma$ by $\gamma(W)$.

On the basic of principle of the grey relation degree, we can know that the greater the grey relation degree $\gamma$, the effecter the comprehensive forecast method.

On the basic of the comprehensive forecast model of the grey relation degree, we can construct the model such as([10]);

$$\max \gamma(W) = \frac{1}{N}\sum_{t=1}^{N} \frac{\min_{1\leq j\leq m}\min_{1\leq t\leq N}|e_j(t)| + \rho \max_{1\leq j\leq m}\max_{1\leq t\leq N}|e_j(t)|}{\left|\sum_{j=1}^{m} w_j e_j(t)\right| + \rho \max_{1\leq j\leq m}\max_{1\leq t\leq N}|e_j(t)|}$$

$$s.t. \begin{cases} \sum_{j=1}^{m} w_j = 1 \\ w_j \geq 0, \quad j=1,2,\cdots,m \end{cases}$$

By solving this optimal problem, we can determine the combinational weight.
$W = \{w_1, w_2, \cdots, w_m\}^T$ based on the optimal grey relation degree.

**5.5. The determination methods of combinational weight based on the theory of rough Set.**

The comprehensive forecast method by statistical rough set theory is to convert the determination problem of weight coefficient to the estimation problem of importance for attribute.

In this technique, first we put on that the set constructed by each sorts of forecast method is the conditional set of decision table.

Then, by using the normalization condition entropy function, we calculate importance degree of decision set (forecast index) for every kind of attribute (forecasting method).

At the final stage, we determine the weight of each kind of forecasting method on the ground of the importance degree. This method can perfectly be analyzed the importance degree of each kind of forecasting method in the data analysis and can perfectly be overcame the subjectivity ([3]) in comprehensive method.

*[ Determination of equivalence class]*

Let U be a universe of discourse and $C$ be a conditional set, $D$ be a decision set.

We denote that column is attribute and row is object (method) in two-dimension data representation system constructed as above.

We can see that each attribute is corresponded to single equivalence relation and whole table define an equivalence relation of unit family.

In order to convenient the classification of knowledge, we progress the specialization treatment for attribute value along some rule.

$U|ind(C) = \{c_1, c_2, \cdots, c_n\}$, $U|ind(D) = \{d_1, d_2, \cdots, d_m\}$

Then, the equivalence class of relation $ind(C)$ and relation $ind(D)$ are called the
conditional class and decision class respectively.

*[ Determination of weight coefficient]*

First, we think that the given decision making method is universal object and then distinguish the quantitative dependence degree of decision set for conditional set.

Then, we eliminate a kind of forecast method in order and calculate the dependence degree at that moment

Lastly, we confirm the importance of knowledge for each kind of forecast method
Thus, the importance of each attribute that is, weight is obtained such as procedure;
Step 1. Calculate the degree of dependence $H(D|C)$ of knowledge $R_D$ for knowledge $R_C$

$$\begin{cases} H(D|C) = \sum_{i=1}^{n} p(c_i) H(D|c_i) \\ H(D|c_i) = -\sum_{j=1}^{n} p(d_i|c_i) \log p(d_i|c_i) \end{cases}$$

Step 2. After eliminating an attribute $a \in C$ in event attribute set, calculate the degree of dependence $H(D|C-\{a\})$ on that occasion by using above expression in accordance with same idea

Step 3. Calculate the degree of importance $\sigma_D(a)$

$$\sigma_D(a) = H(D|(C-\{a\})) - H(D|C)$$

Step 4. Iterating the discussion in this manner, the degree of importance for each attribute can be determined and weight coefficient of $j$'th attribute can be denoted such as;

$$\mu_j = \frac{\sigma_D(j)}{\sum_{i \in c} \sigma_D(i)}$$

**5.6. Comprehensive forecast**

The determination of combination weight in comprehensive forecast by rough set theory has been progressed as following method. That is, the forecast has been progressed for $n$ sort of verification method according to the $k$ sort of forecast method.

We put the forecast results are conditional set for each forecast method the actual data are decision-making set.

Then, we divide data to five sort of class according to magnitude and decide importance weight.

Let $\hat{y}_1(t), \hat{y}_2(t), \cdots, \hat{y}_m(t)$ be predicted values of $m$ sort of another model in time $t$ and $\mu_j\ (j=1,2,\cdots,m)$ be the weight of $j$'th model obtained by the theory of rough set.

Then, the comprehensive forecast model added weight is given by

$$\hat{y}(t) = \sum_{j=1}^{m} \mu_j \hat{y}_j(t), \quad \sum_{j=1}^{m} \mu_j = 1$$

# 6. The comparative analysis of forecasting performance
*[Benchmark of forecasting performance]*

There are many ways to evaluate the forecasting performance of model, ranging from directional measures to magnitude measures to distributional measures.

In practice, the available data set is divided into two sub samples. The first sub sample of the data is used to build a model, and the second sub sample is used to evaluate the forecasting performance of the model.

Let $l$ be forecast length and $y = \{y(1), y(2), \cdots, y(k), \cdots, y(l)\}$ be original sequence to be checked, $\hat{y} = \{\hat{y}(1), \hat{y}(2), \cdots, \hat{y}(k), \cdots, \hat{y}(l)\}$ be sequence obtained by forecast.

And let $\varepsilon = \{\varepsilon(1), \varepsilon(2), \cdots, \varepsilon(N)\}$ be error sequence.

Here $\varepsilon = \{\varepsilon(1), \varepsilon(2), \cdots, \varepsilon(l)\} = \{y(1) - \hat{y}(1),\ y(2) - \hat{y}(2), \cdots,\ y(N) - \hat{y}(l)\}$.

*Magnitude Measure*

Three statistics are commonly used to measure performance of point forecasts.

Those are the mean squared error (MSE), mean absolute error (MAE), and mean absolute percentage error (MAPE).

Mean-square error- MSE and Mean-absolute error- MAE are defined as

$$MSE = \frac{1}{l}\sum_{t=1}^{l}\varepsilon(t)^2 = \frac{1}{l}\sum_{t=1}^{l}[y(t)-\hat{y}(t)]^2$$

$$MAE = \frac{1}{l}\sum_{i=1}^{l}|\varepsilon(i)| = \frac{1}{l}\sum_{i=1}^{l}|y(t)-\hat{y}(t)|$$

Root of mean square error-RMSE and mean absolute percent error-MAPE are defined as

$$RMSE = \sqrt{\frac{1}{l}\sum_{t=1}^{l}[y(t)-\hat{y}(t)]^2}\ ,\quad MAPE = \frac{1}{l}\sum_{t=1}^{l}\left|\frac{y(t)-\hat{y}(t)}{y(t)}\right|\times 100$$

Absolute relative error and mean- absolute relative error are defined as

$$q(k) = \frac{|y(k)|}{|\varepsilon(k)|}\ ,\quad \bar{q} = \frac{1}{n}\sum_{k=1}^{n}q(k)$$

Average forecast accuracies is defined as
$$Accuracy = (1-\bar{q})\times 100\%$$

*[Distributional Measure]*

We have used the benchmark of distributional measure such as;

(ⅰ) The test of ratio of mean square deviations: $C = \dfrac{S_1}{S_2}$

Here $S_1^2 = \dfrac{1}{l}\sum_{k=1}^{l}[\varepsilon(k)-\bar{\varepsilon}(k)]^2$ is mean deviation of error $S_2^2 = \dfrac{1}{l}\sum_{k=1}^{l}[y(k)-\bar{y}(k)]^2$ is mean deviation of original sequence $\bar{\varepsilon} = \dfrac{1}{l}\sum_{k=1}^{l}\varepsilon(k)$; error means, $\bar{y} = \dfrac{1}{l}\sum_{k=1}^{l}y(k)$; mean of original sequence

(ⅱ) Probability of small error: $p = \dfrac{1}{n}P\{|\varepsilon(k)| < 0.6745 S_2\}$

Here $P$ is Probability

(ⅲ) Theil Coefficient: $u^2 = \dfrac{\sum_k [x^{(0)}(k)-\hat{x}^{(0)}(k)]^2 \big/ n}{\sum_k [\hat{x}^{(0)}(k)]^2 \big/ n}$

*[Directional Measure]*

A directional measure considers the future direction (up or down) implied by the model. Predicting that tomorrow's exchange rate of Euro/USD will go up or down is an example of directional forecasts that are of practical interest.

We have used the benchmark of directional measure such as;

$$\text{Directivity (Consistency)} = \frac{1}{l}\sum_{i=1}^{l} d(i) \times 100 \, (\%)$$

Here $d(i) = \begin{cases} 1, & (y(i)-y(i-1))(\hat{y}(i)-y(i-1)) \geq 0 \\ 0, & otherwise \end{cases}$ , $l$ - prediction length,

We have also used benchmark of forecasting performance (Feasibility) such as;

$$\text{Feasibility} = \frac{f_l}{l} \times 100 \quad (\%),$$

Here $f_l = \left|\left\{ i \,\middle|\, |\varepsilon(i)| \leq 0.005, i=1,...,l \right\}\right|$ (%)

We have suggested the Wavelet-SVR-Markov model for forecast of financial time series and compared with the performance of comprehensive forecast model combined by some method of multi-variety for this model. The time series for examination of forecast performance has been chosen exchange rate data- Euro/USD (2007 /1/1~ 2011/11/15).The forecast experiment has been progressed using by MATLAB 7.8 ( $2009a$ ).The seven models have been used to examination of forecast performance.

Those methods are given such as;

The Method 1: The four dimensional vector data composed by open, highest, lowest, close exchange rate have been used to input pattern of SVR and has been forecasted the close exchange rate by Wavelet-SVR –Markov model.

The Method 2: The close exchange rate have been used to input pattern of Wavelet -SVR-Markov and the results of two- SVR model corresponded to each other two parameter has been combined with average weight

The Method 3: The close exchange rate has been forecasted by only Wavelet-SVR- Markov model using close exchange rate data.

The Method 4: The close exchange rate has been forecasted by only SVR model using close exchange rate data.

The Method 5: The close exchange rate have been used to input pattern of Wavelet-SVR-Markov and the results of two- Wavelet -SVR-Markov model corresponded to each other two parameter has been combined with the grey relation degree

The Method 6: The close exchange rate have been used to input pattern of Wavelet-SVR-Markov and the results of two- Wavelet -SVR-Markov model corresponded to each other two parameter has been combined with least square method(optimal combining weight).

The Method 7: The close exchange rate have been used to input pattern of Wavelet-SVR- Markov and the results of two- Wavelet -SVR-Markov model corresponded to each other two parameter has been combined with grey relation optimal combining weight.

Our comprehensive forecasting scheme is given such as;

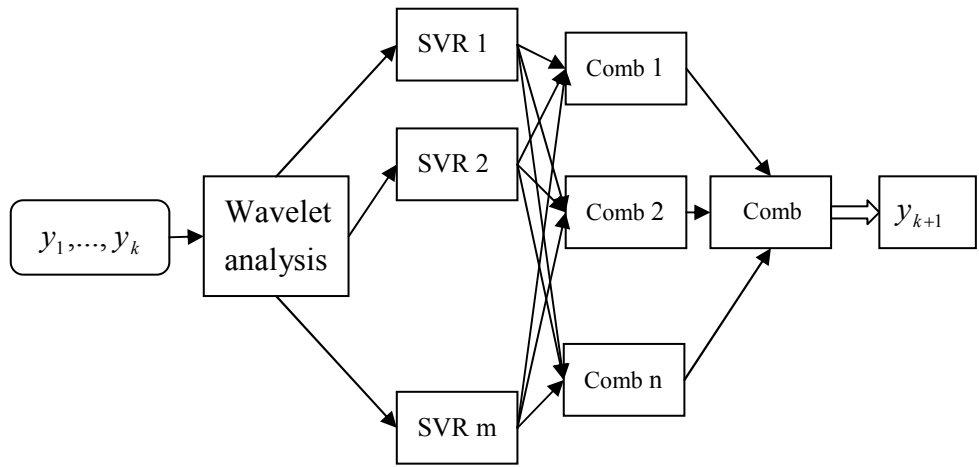

Fig 1.   Two-stage comprehensive forecasting model

*[Combining of two models]*

The data of day's 70 before 31 August. 2011 year is used to build model and after that exchange rate of day's 22 is forecasted by two forecast method.

Then, by using that results, the combining weight has been determined on the ground of five sort of decision method of combining weight that is rough set method( RS-$n$ ,here $n$ is classification number), grey relation optimal combining method GRO, degree of grey coincidence method GRD, least square method LSM, efficient degree method ED.

And then the data of day's 70 before 30 September. 2011 year is used to build model by using five sorts of models corresponding to these weights and after that exchange rate of day's 22 is forecasted. The forecast results are given below table. The numbers in tables are number of forecasting method and letters are name of combining forecast model. In each column of combination weights item of table 1~table 4 the combing weights corresponding to each combining methods are given.

In each row of method item of table 1~table 4 the performance of results forecasted corresponding to each combining methods are given.

Table 1. The forecasting Performance and combining weights--2011/9/30(sample 70, predict 22)

| Method | Performance | | | | | Combination weights | | | | |
|---|---|---|---|---|---|---|---|---|---|---|
| | accuracy | feasibility | consistency | MAE | RMSE | RS-5 | GRO | GRD | LSM | ED |
| 1 | 98.9598 | 18.1818 | 50 | 0.0142 | 0.0197 | 0.3932 | 0 | 0 | 0 | 0.4957 |
| 2 | 99.1413 | 36.3636 | 54.5455 | 0.0118 | 0.0148 | 0.6068 | 1 | 1 | 1 | 0.5043 |
| RS-5 | 99.2048 | 31.8182 | 54.5455 | 0.0109 | 0.0147 | | | | | |
| GRO | 99.1413 | 36.3636 | 54.5455 | 0.0118 | 0.0148 | | | | | |
| GRD | 99.1413 | 36.3636 | 54.5455 | 0.0118 | 0.0148 | | | | | |
| LSM | 99.1413 | 36.3636 | 54.5455 | 0.0118 | 0.0148 | | | | | |
| ED | 99.2078 | 31.8182 | 54.5455 | 0.0108 | 0.0152 | | | | | |

Table 2. The forecasting Performance and combining weights--2011/9/30(sample 70, predict 22)

| Method | Performance | | | | | Combination weights | | | | |
|---|---|---|---|---|---|---|---|---|---|---|
| | accuracy | feasibility | consistency | MAE | RMSE | RS-5 | GRO | GRD | LSM | ED |
| 1 | 98.9598 | 18.1818 | 50 | 0.0142 | 0.0197 | 0.4233 | 0 | 0.4447 | 0.0227 | 0.4958 |

Table 3. The forecasting Performance and combining weights--2011/9/30(sample 70, predict 22)

| Method | Performance | | | | | Combination weights | | | | |
|---|---|---|---|---|---|---|---|---|---|---|
| | accuracy | feasibility | consistency | MAE | RMSE | RS-5 | GRO | GRD | LSM | ED |
| 1 | 98.9598 | 18.1818 | 50 | 0.0142 | 0.0197 | 0.4233 | 0 | 0.4447 | 0.0227 | 0.4958 |
| 4 | 99.0123 | 27.2727 | 59.0909 | 0.0122 | 0.0149 | 0.5767 | 1 | 0.5553 | 0.9773 | 0.5042 |
| RS-5 | 99.1322 | 18.1818 | 54.5455 | 0.0104 | 0.0146 | | | | | |
| GRO | 99.0124 | 27.2727 | 59.0909 | 0.0122 | 0.0149 | | | | | |
| GRD | 99.1383 | 18.1818 | 54.5455 | 0.0104 | 0.0147 | | | | | |
| LSM | 99.0124 | 27.2727 | 59.0909 | 0.0120 | 0.0147 | | | | | |

Table 4. The forecasting Performance and combining weights--2011/9/30(sample 70, predict 22)

| Method | Performance | | | | | Combination weights | | | | |
|---|---|---|---|---|---|---|---|---|---|---|
| | accuracy | feasibility | consistency | MAE | RMSE | RS-5 | GRO | GRD | LSM | ED |
| 1 | 98.8212 | 13.6363 | 50 | 0.0161 | 0.0210 | 0.3673 | 0.3346 | 0.4793 | 0.2949 | 0.4994 |
| 4 | 99.2763 | 40.9091 | 63.6364 | 0.0099 | 0.0137 | 0.6327 | 0.6654 | 0.5207 | 0.7051 | 0.5006 |
| RS-5 | 99.3094 | 40.9091 | 63.6364 | 0.0095 | 0.0125 | | | | | |
| GRO | 99.3174 | 36.3637 | 63.6364 | 0.0094 | 0.0124 | | | | | |
| GRD | 99.2823 | 31.8182 | 63.6364 | 0.0099 | 0.0130 | | | | | |
| LSM | 99.3268 | 31.8182 | 68.1824 | 0.0092 | 0.0124 | | | | | |
| ED | 99.2589 | 31.8182 | 63.6364 | 0.0102 | 0.0133 | | | | | |

The couple of kernel parameter and regulation parameter in method 1, method 3, method 4 are given by [0.0064, 591.9507], [0.1245, 639.559], [0.0241, 687.4275] respectively.

And method 2, method 5~ method 7 are the various combined models of method 3, method 4

*[ Two stage comprehensive forecasting]*

In first stage, we have combined two forecast method by some combining weight decision method as above and progressed forecasting by that respectively.

In two stages, we have again combined the forecast results by the various combined models and progressed forecasting by that .

The several of forecast test results were shown in table 5~ table 8.

These methods have been obtained such as method.

First of all, the results by two sort of method are combined with five sort of combinational methods And then the five sort of forecast method obtained by above method has been combined and obtained the final forecast result.

Table 5. The forecasting Performance and combining weights--2011/9/30

(sample 200, predict 22, method 1,2 )

| Method | Performance | | | | | Combination weights | | | | |
|---|---|---|---|---|---|---|---|---|---|---|
| | accuracy | feasibility | consistency | MAE | RMSE | GRD | LSM | ED | RS-5 | GRO |
| RS-5 | 99.3268 | 31.8182 | 68.1824 | 0.0092 | 0.0124 | 0 | 1 | 0 | 0 | 0 |
| GRD | 99.3 | 40.9091 | 63.6364 | 0.0096 | 0.0126 | 0.1978 | 0.2009 | 0.1956 | 0.2025 | 0.2032 |
| LSM | 99.299 | 40.9091 | 63.6364 | 0.0096 | 0.0126 | 0.2 | 0.2 | 0.2 | 0.2 | 0.2 |
| ED | 99.299 | 40.9091 | 63.6364 | 0.0096 | 0.0126 | 0.2 | 0.2 | 0.2 | 0.2 | 0.2 |
| GRO | 99.3174 | 36.3636 | 63.6364 | 0.0094 | 0.0124 | 0.0265 | 0.3903 | 0.0216 | 0.2478 | 0.3138 |

Table 6. The forecasting Performance and combining weights--2011/9/30

(sample 70, predict 22, method 1,4)

| Method | Performance | | | | | Combination weights | | | | |
|---|---|---|---|---|---|---|---|---|---|---|
| | accuracy | feasibility | consistency | MAE | RMSE | GRD | LSM | ED | RS-5 | GRO |
| RS-7 | 99.141 | 22.7273 | 45.455 | 0.0117 | 0.0158 | 0 | 0 | 1 | 0 | 0 |
| GRD | 99.112 | 31.8182 | 40.909 | 0.0121 | 0.0156 | 0.1931 | 0.2104 | 0.1915 | 0.1947 | 0.2104 |
| LSM | 99.012 | 27.2727 | 40.909 | 0.0135 | 0.0168 | 0.0000 | 0.5 | 0.0000 | 0.0000 | 0.5 |
| ED | 99.111 | 31.8182 | 40.909 | 0.0121 | 0.0156 | 0.1996 | 0.2006 | 0.1995 | 0.1997 | 0.2006 |
| GRO | 99.012 | 27.2727 | 40.909 | 0.0135 | 0.0168 | 0.0000 | 0.5 | 0.0000 | 0.0000 | 0.5 |

Table7. The forecasting Performance and combining weights--2011/9/30

(sample 70, predict 22, method 1,3)

| Method | Performance | | | | | Combination weights | | | | |
|---|---|---|---|---|---|---|---|---|---|---|
| | accuracy | feasibility | consistency | MAE | RMSE | GRD | LSM | ED | RS-5 | GRO |
| RS-8 | 99.1935 | 22.7273 | 63.6364 | 0.0110 | 0.0143 | 0 | 0.5 | 0 | 0 | 0.5 |
| GRD | 99.2487 | 31.8182 | 54.5455 | 0.0103 | 0.0144 | 0.1942 | 0.2084 | 0.1906 | 0.1954 | 0.2114 |
| LSM | 99.1905 | 22.7273 | 63.6364 | 0.0111 | 0.0143 | 0.0000 | 0.5284 | 0.0000 | 0.0259 | 0.4457 |
| ED | 99.2487 | 31.8182 | 54.5455 | 0.0103 | 0.0144 | 0.1997 | 0.2005 | 0.1995 | 0.1998 | 0.2005 |
| GRO | 99.1955 | 27.2727 | 63.6364 | 0.0110 | 0.0143 | 0.0300 | 0.4948 | 0.0287 | 0.0306 | 0.4158 |

Table8. The forecasting Performance and combining weights--2011/9/30

(sample 200, predict 22, method 1,5)

| Method | Performance | | | | | Combination weights | | | | |
|---|---|---|---|---|---|---|---|---|---|---|
| | accuracy | feasibility | consistency | MAE | RMSE | GRD | LSM | ED | RS-5 | GRO |
| RS-9 | 99.2703 | 31.8182 | 63.6364 | 0.0100 | 0.0131 | 1 | 0 | 0 | 0 | 0 |
| GRD | 99.3032 | 40.9091 | 63.6364 | 0.0095 | 0.0126 | 0.1978 | 0.2009 | 0.1956 | 0.2025 | 0.2032 |
| LSM | 99.3029 | 40.9091 | 63.6364 | 0.0095 | 0.0126 | 0.2 | 0.2 | 0.2 | 0.2 | 0.2 |
| ED | 99.3029 | 40.9091 | 63.6364 | 0.0095 | 0.0126 | 0.2 | 0.2 | 0.2 | 0.2 | 0.2 |
| GRO | 99.3172 | 36.3636 | 63.6364 | 0.0094 | 0.0124 | 0.0266 | 0.3903 | 0.0217 | 0.2478 | 0.3136 |

*[Result analysis]*

First, we have compared with the results obtained by various methods for the forecast results of two SVR model correspond to each other different parameter couple. Those results are given at table 9. Here $h$ as known from table 9, when two sort of SVR model has been combined, the method 5 that is combinational method by degree of grey coincidence is the best for all indicator of forecast performance.

These facts can clearly know through the rank by GCTOPSIS (R1), PROTOPSIS (R2) and RTOPSIS (R3), respectively, and the final rank (R).

Where GCTOPSIS is grey correlation TOPSIS, PROTOPSIS is projected weight grey relation TOPSIS and RTOPSIS is distance-based TOPSIS.

In the forecast of 54 day, 77 day, 120 day, the comprehensive rank of the 5 method has been occupied all the first place for these ranks and the 2 method was the second place

Proceeding the comprehensive estimation, we have such that result : method 5 > method 2 > method 6 > method 7 .

That is, we have such that result : degree of grey coincidence method > average weight method > least square weight method > optimal grey coincidence combining weight method

Let us now progress to the estimation of forecasting results by method of decision-making plan estimation for multiple attributive decision-making methods

We put relative accuracy, feasibility, consistency, mean absolute error -MAE, root of mean-squared error-RMSE which take part in the forecast performance indicator are estimation indicators and the forecasting methods are the decision -making plan in decision –making method..

Then, we take subjective weights of elative accuracy, feasibility, consistency, mean absolute error, root of mean-squared error with (0.15, 0.2, 0.3, 0.2, 0.15)

When comparing the results of table 4 and table 5, table 3and table 6, table 2 and table 7, we can know that secondary comprehensive result is better than first comprehensive result.

Then, we have ranged the results of first comprehensive model according to the results of table 1~table 4.

The ranged results are given in table 10.Here the numbers of comprehensive models are given such as: 1-RS; 2-GRO; 3-GRD; 4-LSM; 5-ED.

Then, we have ranged the results of secondary comprehensive model according to the results of table 5~table 8.The ranged results are given in table 11. Here the numbers of comprehensive models are given such as: 1-RS; 2-GRD; 3-LSM; 4-ED; 5-GRO.From table 10 and table 11, when have been progressed the first comprehensive model, we can know that arrangement rank is given such as; Grey relation optimal comprehensive method>Rough set method> Grey relation degree method> Least square method> Efficient degree method.When have been progressed the secondary comprehensive model, we can know that arrangement rank is given such as; Grey relation degree method> Efficient degree method> Least square method> Rough set method> Grey relation optimal comprehensive method. Combining two ranks, we obtain final range such as: Grey relation degree method> Rough set method> Grey relation optimal comprehensive method>Efficient degree method> Least square method.

Table9. The forecasting Performance various models and rank --2011/9/30-
(sample 200, predict 22, 54,77, 98,120)--

| Method | Performance | | | | | h | Rank | | | |
|---|---|---|---|---|---|---|---|---|---|---|
| | accuracy | feasibility | consistency | MAE | RMSE | | R1 | R2 | R3 | R |
| 2 | 99.111 | 31.818 | 59.091 | 0.0122 | 0.0147 | 22 | 2 | 2 | 3 | 2 |
| 5 | 99.113 | 27.273 | 54.545 | 0.0122 | 0.0147 | | 3 | 3 | 4 | 3 |
| 6 | 99.126 | 36.364 | 59.091 | 0.0120 | 0.0147 | | 1 | 1 | 1 | 1 |
| 7 | 99.13 | 22.727 | 54.545 | 0.0121 | 0.0141 | | 4 | 4 | 2 | 4 |
| 2 | 99.204 | 33.333 | 66.667 | 0.0109 | 0.0140 | 54 | 2 | 2 | 3 | 2 |
| 5 | 99.205 | 35.185 | 66.667 | 0.0109 | 0.0140 | | 1 | 1 | 1 | 1 |
| 6 | 99.202 | 31.481 | 64.815 | 0.0109 | 0.0140 | | 3 | 3 | 4 | 3 |
| 7 | 99.225 | 31.481 | 59.259 | 0.0106 | 0.0140 | | 4 | 4 | 2 | 4 |
| 2 | 99.259 | 35.065 | 67.532 | 0.0103 | 0.0131 | 77 | 2 | 2 | 2 | 2 |
| 5 | 99.263 | 35.065 | 68.831 | 0.0102 | 0.0131 | | 1 | 1 | 1 | 1 |
| 6 | 99.258 | 33.766 | 66.234 | 0.0103 | 0.0131 | | 3 | 3 | 3 | 3 |
| 7 | 99.264 | 32.468 | 63.636 | 0.0102 | 0.0134 | | 4 | 4 | 4 | 4 |
| 2 | 99.287 | 34.694 | 65.306 | 0.0099 | 0.0126 | 98 | 2 | 2 | 2 | 2 |
| 5 | 99.289 | 34.694 | 66.327 | 0.0099 | 0.0126 | | 1 | 1 | 1 | 1 |
| 6 | 99.278 | 34.694 | 65.306 | 0.0101 | 0.0127 | | 3 | 3 | 4 | 3 |
| 7 | 99.299 | 33.673 | 65.306 | 0.0098 | 0.0129 | | 4 | 4 | 3 | 4 |
| 2 | 99.287 | 32.5 | 64.167 | 0.0099 | 0.0125 | 120 | 2 | 2 | 3 | 2 |
| 5 | 99.289 | 32.5 | 65 | 0.0099 | 0.0125 | | 1 | 1 | 1 | 1 |
| 6 | 99.28 | 31.667 | 64.167 | 0.0101 | 0.0126 | | 4 | 4 | 4 | 4 |
| 7 | 99.298 | 31.667 | 65 | 0.0098 | 0.0127 | | 3 | 3 | 2 | 3 |

Table 10.

| Table | Rank | | | | | Ranking method | Comprehensive rank | Final rank |
|---|---|---|---|---|---|---|---|---|
| 1 | 2 | 3 | 4 | 5 | 1 | GCTOPSIS | 2 3 4 1 5 | 2 1 3 4 5 |
| | 2 | 3 | 4 | 5 | 1 | PROTOPSIS | | |
| | 1 | 2 | 3 | 4 | 5 | RTOPSIS | | |
| 2 | 3 | 1 | 5 | 4 | 2 | GCTOPSIS | 3 1 4 5 2 | |
| | 3 | 1 | 5 | 4 | 2 | PROTOPSIS | | |
| | 4 | 2 | 1 | 3 | 5 | RTOPSIS | | |
| 3 | 2 | 4 | 5 | 1 | 3 | GCTOPSIS | 2 5 4 1 3 | |
| | 2 | 4 | 5 | 1 | 3 | PROTOPSIS | | |
| | 5 | 1 | 3 | 2 | 4 | RTOPSIS | | |
| 4 | 1 | 2 | 4 | 3 | 5 | GCTOPSIS | 1 2 4 3 5 | |
| | 1 | 2 | 4 | 3 | 5 | PROTOPSIS | | |
| | 4 | 1 | 2 | 3 | 5 | RTOPSIS | | |

Table 11.

| Table | Rank | | | | | Ranking method | Comprehensive rank | Final rank |
|---|---|---|---|---|---|---|---|---|
| 5 | 2 | 3 | 4 | 5 | 1 | GCTOPSIS | 2  3  5  1  4 | |
|   | 2 | 3 | 4 | 5 | 1 | PROTOPSIS | | |
|   | 1 | 5 | 2 | 3 | 4 | RTOPSIS | | |
| 6 | 2 | 4 | 3 | 5 | 1 | GCTOPSIS | 2  4  1  3  5 | |
|   | 2 | 4 | 1 | 3 | 5 | PROTOPSIS | | 2  4  3  1  5 |
|   | 1 | 2 | 4 | 3 | 5 | RTOPSIS | | |
| 7 | 2 | 4 | 5 | 1 | 3 | GCTOPSIS | 2  4  5  1  3 | |
|   | 2 | 4 | 5 | 1 | 3 | PROTOPSIS | | |
|   | 5 | 2 | 4 | 1 | 3 | RTOPSIS | | |
| 8 | 2 | 3 | 4 | 5 | 1 | GCTOPSIS | 2  3  4  1  5 | |
|   | 2 | 3 | 4 | 5 | 1 | PROTOPSIS | | |
|   | 1 | 2 | 3 | 4 | 5 | RTOPSIS | | |

## 7. Conclusion

In this paper, we have suggested a Wavelet-SVR-Markov forecasting method to forecast financial time series and showed that the forecast performance can be more improved with rational combinations through various combinational experiments. Through our experiment results have been showed that two stage combinational model is finer than general combinational model.

We also have estimated synthetically the combinational forecast methods according to various forecast performance indicator. Through those results have been showed that grey relation degree method and grey relation optimal comprehensive method has good forecast performance.